\documentclass[12pt,epsf,showkeys,nofootinbib]{article}

\usepackage{graphicx,float}
\usepackage{mathrsfs,array,multirow}
\usepackage{amstext}
\usepackage{subfigure}
\usepackage{bm,latexsym,amsmath,amsfonts,amssymb}
\usepackage[usenames,dvipsnames]{color}
\usepackage{color}
\usepackage[colorlinks=true,linkcolor=blue]{hyperref}
\usepackage{soul}
\usepackage{epsfig}

\newcommand{\be}[1]{\begin{equation} \label{#1}}
\newcommand{\ee}{\end{equation}}

\newcommand{\ba}{\begin{array}}
\newcommand{\ea}{\end{array}}

\newcommand{\bea}{\begin{eqnarray}}
\newcommand{\eea}{\end{eqnarray}}

\begin{document}
\begin{titlepage}
\vspace{.5in}
\begin{flushright}
CQUeST-2025-0755
\end{flushright}
\vspace{0.5cm}

\begin{center}
{\Large\bf Dressing rotating black holes with anisotropic matter }\\
\vspace{.4in}

  {$\mbox{Hyeong-Chan\,\, Kim}^{\P}$}\footnote{\it email: hckim@ut.ac.kr},\,\,
  {$\mbox{Wonwoo \,\, Lee}^{\S}$}\footnote{\it email: warrior@sogang.ac.kr} \\

\vspace{.3in}

{\small \P \it School of Liberal Arts and Sciences, Korea National University of Transportation, Chungju 27469, Korea}\\
{\small \S \it Center for Quantum Spacetime, Sogang University, Seoul 04107, Korea}\\

\vspace{.5in}
\end{center}
\begin{center}
{\large\bf Abstract}
\end{center}
\begin{center}
\begin{minipage}{4.75in}

{\small \,\,\,\, We present a new rotating black hole solution to the Einstein equations as an extension of the Kerr spacetime.
Interestingly, the solution we find may not be uniquely characterized by asymptotic parameters such as mass,
angular momentum, and charge, thereby it would be the additional hair.
We also analyze in detail how this additional characteristics or this hair affects the thermodynamic properties of the black hole.

 }
\end{minipage}
\end{center}
\end{titlepage}

\newpage

\section{Introduction \label{sec1}}

\quad

The study of black holes has made significant progress, particularly following their indirect detections~\cite{Akiyama:2019cqa, Akiyama:2019bqs, Akiyama:2019fyp, TheLIGOScientific:2016src, Abbott:2016nmj, Abbott:2017vtc, Bae:2023sww} not only provide a deeper understanding of various astrophysical phenomena but also serve as an intriguing subject of theoretical investigation.
In particular, static, spherically symmetric black holes offer an idealized framework for analyzing complex gravitational systems.
Once these simpler models are well understood, the insights gained can pave the way for studying rotating black holes.
However, this transition poses considerable challenges due to the increased complexity of the equations governing their geometry. The computational demands of studying rotating black holes often necessitate simplifications, such as assuming axial symmetry or focusing on static cases.

A key solution describing a rotating black hole in vacuum is the Kerr metric~\cite{Kerr:1963ud}, which serves as a foundational reference for deriving more general rotating black hole solutions that incorporate matter fields.
The Kerr solution provides a remarkably accurate description of astrophysical black holes and is widely accepted as the best approximation for rotating black holes in the Universe. However, real black holes in galaxies exist in environments filled with dark matter and dark energy.
Thus, it is necessary to explore more general solutions that take into account the effects of a cosmological constant and matter fields.
Studying such extensions can offer deeper insights into the dynamics and observational signatures of astrophysical black holes.

This naturally raises the question: How can we derive solutions for rotating black holes that coexist with surrounding matter fields?
The first major extension in this direction was the Kerr-Newman solution~\cite{Newman:1965my}, which describes a rotating, charged black hole in Einstein-Maxwell theory.
This solution was derived using the Newman-Janis (NJ) algorithm~\cite{Newman:1965tw}, a method that has been widely studied and debated~\cite{Drake:1998gf, Azreg-Ainou:2014aqa, Azreg-Ainou:2014pra, Adamo:2014baa, Beltracchi:2021ris, Kim:2024dxo, Kim:2024mpy}.
While the underlying reason how the algorithm works remains unclear, its empirical success makes it a valuable tool for deriving rotating black hole solutions.
Recent research has focused on extending this algorithm to incorporate additional matter fields~\cite{Toshmatov:2015npp, Xu:2016jod, Kim:2019hfp, Kim:2021vlk}.
There is also considerable interest in which types of fields can coexist with rotating black holes while maintaining physical consistency~\cite{Janis:1965tx, Erbin:2014aya, Kim:2021vlk}.
In various theories of gravity, efforts have been made to introduce new rotating black hole solutions in different spacetime dimensions~\cite{Taub:1981evj, Dereli:1986cm, Sen:1992ua, Kim:1998hc, Gibbons:2004js, Bambi:2013ufa, Herdeiro:2014goa, Brahma:2020eos, Simpson:2021dyo, Masa:2022weh, Devecioglu:2024uyi, Nurmagambetov:2024hsn, AraujoFilho:2024rss, Ghosh:2025igz, Li:2025glq}.

To study the matter fields coexisting with a rotating black hole,
it is crucial to analyze the components of the stress-energy tensor
and understand the equation of state, which determines the nature of the matter~\cite{Stephani:2003tm}.
Such an analysis provides valuable insights into the behavior of the matter fields surrounding the rotating black holes.
For the electrically charged Kerr-Newman black hole, the stress-energy tensor reveals that
the energy density is related to the radial pressure by the equation of state $\varepsilon= -p_r$.
Interestingly, Maxwell's field having this property does not behave as a perfect fluid.
When considering more general matter fields, two main approaches emerge:
First is to find a rotating black hole solution that coexists with matter of the perfect fluid type~\cite{Delgaty:1998uy, Semiz:2008ny}.
The second involves extending the Kerr-(Newman) black hole to
the rotating black hole with matter fields having the property $\varepsilon= -p_r$, such as electric fields~\cite{Ruderman:1972aj, Herrera:1982, Herrera:1997plx, Bowers:1974tgi, Matese:1980zz, Mak:2001eb, Kiselev:2002dx, Thirukkanesh:2008xc, Ivanov:2002xf, Varela:2010mf, Cho:2017nhx, Kim:2019ygw, Toshmatov:2015npp, Xu:2016jod, Kim:2019hfp, Kim:2021vlk}.

The no-hair theorem is well-established for the electro-vacuum solution of Einstein-Maxwell theory~\cite{Ruffini:1971bza}, which states that black holes are characterized solely by their mass, charge, and angular momentum as primary hairs. However, in Einstein-Maxwell theory with additional matter fields beyond ordinary matter or in extensions of general relativity, the applicability of the no-hair theorem is not necessarily evident. When fields other than the electromagnetic field are present, additional hairs may emerge.
In the Einstein-scalar theory with minimal coupling, Bekenstein's no-hair theorem holds under the assumption that the energy density of the scalar field outside the black hole horizon is non-negative~\cite{Bekenstein:1972ny, Bekenstein:1995un}. Thus, the theorem can be circumvented if this assumption is violated~\cite{Herdeiro:2015waa, Antoniou:2017acq, Doneva:2017bvd, Lee:2018zym, Minamitsuji:2018vuw, Zou:2020rlv, Doneva:2020qww, Papageorgiou:2022umj, Lee:2021uis, Ghosh:2023kge, Chew:2024rin}. Further investigation is required to determine whether such hair should be classified as primary or secondary~\cite{Campbell:1990ai, Coleman:1991jf, Coleman:1991ku}.

Rather than following the conventional classification, in this work, we categorize black hole hairs in general relativity into two types based on their effect on the geometry, a distinction that is useful for our later discussion.
The first type of hair directly modifies the metric, meaning its presence alters the stress-energy tensor.
This category includes conventional hairs such as mass, charge, and angular momentum, as well as most non-Abelian and dilatonic hairs.
The second type of hair, in contrast, does not affect the metric but can still be detected through global measurements, such as Aharonov-Bohm-type scattering~\cite{Frolov:1998wf}.
Examples include axionic quantum hair and discrete gauge charges.
Interestingly, stress-energy tensors associated with the first type of hair typically decay polynomially at large distances,
making their influence detectable by asymptotic observers.
This observation raises an intriguing question: Could there exist a new type of black hole hair
that remains undetectable to an asymptotic observer yet can be identified through local measurements?
For such a hair to exist, its associated stress-energy tensor must be bounded near the event horizon
and decay no slower than exponentially at large distances.

In this work, we use results of the ``non-complexification procedure''~\cite{Azreg-Ainou:2014pra}, a developments of the Newman-Janis algorithm, as a method for separation of variables for the Einstein equation to explore the structure of axially symmetric, rotating solutions.
Let us briefly summarize the strategy to get the rotating solution in this scheme.
As the base geometry, we consider a static, spherically symmetric metric of the form,
\begin{equation}
ds^2 = - f(r) dt^2 + \frac{dr^2}{g(r)}  + \sigma(r) d\Omega^2 \, ,
\label{metric:wh31}
\end{equation}
where $d\Omega^2 = d\theta^2 + \sin^2\theta d\phi^2$ represents the angular part of the metric.
With black hole solutions in mind, we set $\sigma(r) = r^2$.
At this stage, we do not presume any specific forms for the functions $f(r)$ and $g(r)$ aside from the asymptotic flatness condition, which requires $f(r)\,, g(r) \to 1$  as $r\to \infty$.

The resulting rotating geometries can be written as the form~\cite{Azreg-Ainou:2014pra}
\begin{eqnarray}
\label{metric:ortho gen}
ds^2 &=&
 -  \frac{  \Sigma \Delta}{ \left(\Gamma -   a^2\sin^2\theta \right)^2 } \left( dt - a \sin^2\theta d\phi \right)^2
+ \Sigma d\theta^2 +  \frac{\Sigma}{\Delta }dr^2 \nonumber \\
&&
 + \frac{  \Sigma \sin^2\theta }{  \left(\Gamma -  a^2\sin^2\theta \right)^2 }	(a dt - \Gamma d\phi)^2 \,,
\end{eqnarray}
where $\Sigma(r, \theta)$, $\Delta(r)$, and $\Gamma(r)$ will be shown in Sec.~\ref{sec2}.
By writing the metric function in this way, the orthonormal basis is easy to read.
The components of the energy-momentum tensor are diagonalized on an orthonormal frame~\cite{Carter:1968ks, Azreg-Ainou:2014nra}.

In this metric, there are three unknown functions to be determined.
Of these, only $\Sigma(r,\theta)$, which generalizes $\sigma(r)$, depends both on $r$ and $\theta$.
The coordinate transformation functions $\Gamma(r)$ and $\Delta(r)$ depend only on $r$.
The physical context should specify the appropriate choices for these functions.
The functions are determined through physical requirements and the Einstein equations with the constraints  in the non-rotating limit,
\begin{equation}
g(r)  \equiv 
  \lim_{a \to 0} \frac{ \Delta(r)}{\Sigma(r,\theta)}
		, \qquad
f(r) \equiv 
	  \lim_{a\to 0} \frac{ \Delta(r) \Sigma(r,\theta) }{ \Gamma^2} \,.
\label{gf:non-rot}
\end{equation}
Noting $\Sigma(r,\theta) \to \sigma(r)$ in the $a \to 0$ limit, this relation constrains the functional forms for $\Delta(r)$ and $\Gamma(r)$.

In this work, we regard the metric~\eqref{metric:ortho gen} as the starting one to analyze the Einstein equations
and get the orthonormal frame.
Therefore, we do not assume the predefined static limit~\eqref{gf:non-rot}
but try to get the solution from the general physical requirements.
We investigate the structure of these solutions and discuss their implications for black hole physics.

The paper is organized as follows:
In Sec.~\ref{sec2}, we derive a new rotating solution of the Einstein equation, which is a generalization of the Kerr black hole coexisting with a matter field.
Sec.~\ref{sec3} is dedicated to analyzing the properties of this new solution in depth.
In Sec.~\ref{sec4},  we explore the implications of the new solution for black hole thermodynamics, with a focus on how the presence of the new matter field affects thermodynamic properties. Finally, we summarize our findings and discuss their broader implications in the concluding section.

\section{Rotating geometries from the static spherically symmetric black hole \label{sec2}}

\quad
We develop {\it new} rotating solutions, which are a generalization of the Kerr geometry.
We aim to find new solutions by modifying the Kerr geometry minimally.
Thus, we follow the same form as the Kerr's for $\Sigma(r,\theta)$:
\begin{equation}
\Sigma(r,\theta) = r^2 + a^2 \cos^2\theta .
\end{equation}
The metric functions are determined from $\Delta(r)$ and $\Gamma(r)$ as given in Eq.~\eqref{metric:ortho gen}, which are prescribed through Einstein equation. Therefore, we choose the following general form:
\begin{equation}
\label{prescription1}
\Delta(r) = r^2+a^2 -2Mr +  v(r) \,, \qquad
\Gamma(r) \equiv r^2 +\tilde A(r) +a^2 \,,
\end{equation}
where $v(r)$ and $\tilde A(r)$ denote the deviations from the Kerr geometry.
The metric~\eqref{metric:ortho gen}  becomes
\begin{eqnarray}
\label{generalized Kerr1}
ds^2 &=& -  \frac{  \Sigma(r,\theta) \Delta(r)}{\left( \Sigma(r,\theta)+ \tilde A(r)\right)^2 }
		\left(dt - a \sin^2\theta \, d\phi\right)^2 + \frac{\Sigma(r,\theta)}{\Delta(r)} dr^2 + \Sigma(r,\theta) d\theta^2 \nonumber \\
&&+\frac{\Sigma(r,\theta) \sin^2\theta}{ \left( \Sigma(r,\theta)+ \tilde A(r)\right)^2 }\left(a dt- \Gamma(r) d\phi \right)^2 \,.
\end{eqnarray}

Then, we require the condition that the stress tensor does not have
the off-diagonal\footnote{Note however that this non-vanishing value does not imply the existence of shear viscosity between the two directions.
One could find the orthonormal basis which makes this value vanish by rotating $r$ and $\theta$ directions.} $r\theta$ component, $T_{ r \theta} =0$.
This is required for fluids undergoing only a rotational motion about a fixed axis (the $z$-axis here), which leads to $R_{ r \theta}=0$~\cite{Azreg-Ainou:2014nra}.
To calculate the Einstein tensor, we take the orthonormal non-coordinate basis from the metric~\eqref{generalized Kerr1}:
\begin{eqnarray}
\label{basis:nc1}
&& \omega^{\hat t} = \frac{  \sqrt{\Sigma(r,\theta) \Delta(r)}}
	{\Sigma(r,\theta)+ \tilde A(r)  }
\left[  dt - a \sin^2\theta d\phi\right], \quad
\omega^{\hat r} = \sqrt{\frac{\Sigma(r,\theta)}{\Delta(r)}} dr \,, \nonumber \\
&& \omega^{\hat \theta} = \sqrt{\Sigma(r,\theta)} d\theta, \quad
\omega^{\hat \phi} = \frac{\sqrt{\Sigma(r,\theta)} \sin\theta}{\Sigma(r,\theta) + \tilde A(r)}
	\left[a dt-  \Gamma(r) d\phi \right] \,.
\end{eqnarray}
Starting from this basis, we construct the connection 1-form through $d\omega^{\hat a}= -\omega^{\hat a}_{~\hat b} \wedge \omega^{\hat b}$ and find the curvature 2-form, $\mathfrak{R}^{\hat a}_{~\hat b}=d\omega^{\hat a}_{~\hat b}+\omega^{\hat a}_{~\hat c} \wedge \omega^{\hat c}_{~\hat b}.$
From the coefficients of this curvature 2-forms, we read off the curvature components.
The result is given by
\begin{equation}
R_{\hat r \hat \theta}
=\frac{3 a^2 \sin \theta \cos\theta \sqrt{\Delta(r)} \left(\Sigma^2(r, \theta) \tilde A'(r)
		-4 r \Sigma(r, \theta) \tilde A(r)-2  r \tilde A^2(r)\right)}
   {\Sigma^3(r, \theta)  \left( \Sigma(r, \theta)  + \tilde A(r)\right)^2} \,.
\end{equation}
Through the Einstein equation $G_{ab} = 8\pi T_{ab}$, the absence of the off-diagonal components gives $R_{\hat r \hat \theta}=0$, which leads
\begin{equation}
\label{tilde A}
\tilde A(r) =0   \quad \to \quad \Gamma(r) = r^2 + a^2 \,.
\end{equation}

Taking this value, the metric now becomes
\begin{eqnarray}
\label{metric:gen Kerr}
ds^2 &=& -  \frac{ \Delta(r)}{ \Sigma(r, \theta) } \left(dt - a \sin^2\theta \, d\phi\right)^2
+ \frac{\Sigma(r, \theta)}{\Delta(r)} dr^2 + \Sigma(r, \theta) d\theta^2 \nonumber \\
&& +\frac{\sin^2\theta}{\Sigma(r, \theta)}\left[ a dt-(r^2+a^2) d\phi \right]^2 \,,
\end{eqnarray}
where $\Delta(r) = r^2+a^2-2Mr + v(r)$ as shown in Refs.~\cite{Gurses:1975vu, Toshmatov:2015npp, Kim:2021vlk}.
In Ref.~\cite{Gurses:1975vu}, the Kerr-Schild coordinate was analyzed in detail.
Notice that the metric has almost the same form as the Kerr one except that the function $\Delta(r) $ now includes a correction term, $v(r)$.

Still, we cannot say that the metric~\eqref{metric:gen Kerr} to describe a solution of the Einstein equation.
To ensure the metric to be a solution, one should prescribe what matter make the geometry.
For this purpose, we check the Einstein tensor and check the equation of state required to specify a solution.
We take the tetrad frame~\cite{Carter:1968ks, Azreg-Ainou:2014nra, Kim:2021vlk}
\begin{eqnarray}
\label{otetrad}
e^{\mu}_{\hat{t}}&=& \frac{(r^2+a^2,0,0,a)}{\sqrt{\Sigma \triangle}}\,,~~~~e^{\mu}_{\hat{r}}= \frac{\sqrt{\triangle}(0,1,0,0)}{\sqrt\Sigma} \,, \nonumber \\
e^{\mu}_{\hat{\theta}}&=&\frac{(0,0,1,0)}{\sqrt\Sigma}\,, \quad e^{\mu}_{\hat{\phi}} =-\frac{(a\sin^2\theta,0,0,1)}{\sqrt{\Sigma}\sin\theta} \,.
\end{eqnarray}
The Einstein tensor for this geometry with respect to the basis~\eqref{otetrad} with $\tilde A(r) =0$ takes the diagonal form:
\begin{equation}
G_{\hat t \hat t} = - G_{\hat r \hat r} = \frac{v(r)- rv'(r)}{\Sigma^2(r, \theta)} , \qquad
G_{\hat \theta \hat \theta} =  G_{\hat\phi \hat\phi}
	= \frac{ \Sigma(r, \theta) v''(r)/2+v(r) - r v'(r) }{\Sigma^2(r, \theta)} \,.
\label{gen K}
\end{equation}
Now, let us display a few important cases of the metric.
As seen from the value of the Einstein tensor, the geometry is no longer a vacuum solution  except for specific choices of $v$.
Note, however, that the developed Einstein tensor vanishes when $v\to 0$.
Also, the equation of state for the radial part $w_r \equiv p_r/\rho =-1$, is  always satisfied, which resembles that of the electro-magnetic field for the Kerr-Newman solution.
This is satisfied when $f(r)=g(r)$, in which  the radial pressure could be the negative of the  energy density~\cite{Jacobson:2007tj}.

Note that the two limits $a\to 0$ and $v(r)\to 0$ are independent.
Therefore we can take the $a\to 0$ limit with $v(r)\neq 0$.
In this case, the metric~\eqref{metric:gen Kerr} provides a new static geometry:
\begin{equation}
ds^2 = -  \frac{   \Delta^s(r)}{ r^2 } dt ^2
+ \frac{r^2}{\Delta^s(r)} dr^2 + r^2 d\Omega_{(2)}^2 \,,
\label{gen Sch}
\end{equation}
where $\Delta^s = r^2-2Mr + v(r)$.
The source developing this metric satisfies $T_{\mu\nu} = G_{\mu\nu}/8\pi$ with the Einstein tensor given in Eq.~\eqref{gen K}.
Therefore, we interpret that the metric~\eqref{metric:gen Kerr} describes the rotating geometry of this static one.

When we choose
\begin{equation}
v(r) = c_1 r
	+ K r^{2-2w}  \,,
\label{static1}
\end{equation}
the metric describes a static black hole made of matter with equation of state $w_{\theta} = w_\phi = w$ with $w_r =-1$ known in Ref.~\cite{Cho:2017nhx}.
Since the $c_1$ term represents mass rescaling, we set $c_1 =0$.
Now,
\begin{equation}
v(r) - r v'(r) = K (2w-1) r^{2-2w} \,. \nonumber
\end{equation}
Therefore the positive energy condition holds when
\begin{equation}
K (2w-1) > 0 \, . \nonumber
\end{equation}

Let us display a few properties of the solution~\eqref{metric:gen Kerr} for various special cases.
\begin{itemize}
\item When $v(r)= r v'(r)$, i.e., $v(r) = v_1r$ is linear in $r$, the Einstein tensor vanishes.
The solution describes the Kerr solution with modified mass $M' = M+ v_1/2$.

\item When $v(r) = v_0 + v_1 r$, the equation of state takes the form, $w_1 = -1$, $w_2=w_3=1$, which is the same as that of the electro-magnetic field.
Therefore, the geometry takes the same form as the Kerr-Newman solution with mass $M' = M+  v_1/2$ and the charge squared, $ Q^2 =  v_0$.

\item When $v(r)= v_2 r^2$, the equation of state becomes $w_1 = -1$ and $w_2=w_3 = -a^2\sin^2\theta/r^2$.
The energy density $\rho = T^{\hat t \hat t} = G^{\hat t \hat t}/8\pi = (- v_2 r^2)/\Sigma^2(r, \theta) $.
To interpret this solution, we first consider the $a=0$ case.
Noting the metric~\eqref{gen Sch}, we find that the static metric reproduces the global monopole spacetime.
Therefore, we can conclude that the solutions with $a \neq 0$ is the rotating global monopole~\cite{TeixeiraFilho:2001fc}.

\item When $v(r)$ is given by Eq.~\eqref{static1}, the metric~\eqref{metric:gen Kerr} describes the rotating geometry of the static solution~\eqref{gen Sch}.
The equation of state in this case is $w_r = -1$ and $w_\theta=w_\phi = (w-1)\Sigma(r, \theta)/r^2 +1$.
When $w=1$, the solution correspond to the Kerr-Newman solution.
Note that, from the form of the Einstein tensor~\eqref{gen K}, the angle dependence of the equation of state is unavoidable unless except for the above three cases.

\item
Any functional dependence for $v$ can be constructed by imposing the stress tensor correspondingly once it satisfies the physically acceptable conditions such as asymptotic flatness.
However, to accept as an interesting generalization of Kerr family,
the geometry should be constructed from a minor modification of the equation of state for matter.
A simplest unnotices equation of state $w_1 = -1$, and $w_2=w_3 = 1 + v_c \Sigma(r, \theta)/2 $ is achieved when $v''(r) = v_c (v(r)-rv'(r))$.
In this case,
\begin{equation}
v(r) = v_1 r + v_2 \left[ e^{-v_c r^2/2}
	+ \sqrt{\frac{\pi v_cr^2}2} \, \left( \mbox{erf} \left( \sqrt{\frac{v_cr^2}{2}} \right) -1\right) \right]\,,
\label{v sol}
\end{equation}
where erf$(x)$ denotes the error function.
This is an interesting generalization of the Kerr family with a minor modification of the equation of state.
Note that $v(r)-rv'(r) =  v_2 e^{- v_c r^2/2}$.
Therefore, the stress tensor exponentially decreases with $r$.
We analyze the properties of the solution in the next section.

\item Else, the equation of state for the radial part satisfies $w_1 = -1$.
But, $w_2=w_3$ has an angle dependence.
The geometry still describes a stationary rotating black hole with surrounding matter.
An interesting observation here is that because the stress tensor is linear in $v$, any combinations of the above cases also belongs to a solution of the Einstein equation.
For example, one may add $v_1= q^2$ [electric field], $v_2= K r^{-2}$ [$w=2$ case in Eq.~\eqref{static1}], and $v_3$ [the above erf functions].
\end{itemize}
Starting from the metric ansatz~\eqref{metric:gen Kerr}, we impose the axially rotating condition $R_{\hat r \hat \theta} =0$ and select an appropriate equation of state to determine the final form of the metric.

\section{Analysis of the new solution \label{sec3}}
In this section, we analyze the properties of the newly found generalized Kerr solution~\eqref{v sol} with metric,
\begin{equation}
ds^2 = -  \frac{  \Delta(r)}
	{\Sigma(r, \theta) }
		\left(dt - a \sin^2\theta \, d\phi\right)^2
+ \frac{\Sigma(r, \theta)}{\Delta(r)} dr^2 + \Sigma(r, \theta) d\theta^2
+\frac{\sin^2\theta}{ \Sigma(r, \theta) }\left( \Gamma(r) d\phi - a dt\right)^2 \,.
\label{generalized Kerr}
\end{equation}
 where
\begin{equation}
\Sigma(r, \theta) = r^2 + a^2 \cos^2\theta\,, \qquad
\Delta(r) = r^2+a^2 -2Mr +  v(r)\,, \qquad
\Gamma(r) \equiv r^2+a^2 \,,
\label{prescription11}
\end{equation}
with $v(r)$ given in Eq.~\eqref{v sol}.

The energy density of the new solution~\eqref{v sol} is
\begin{equation}
\label{rho:new}
\rho = T^{\hat{t}\hat{t}} =-T^{\hat{r}\hat{r}} = \frac{ (v(r)-rv'(r))}{8\pi \Sigma^2(r, \theta)}  = \frac{v_2 e^{-v_c r^2/2} }{8\pi \Sigma^2(r, \theta)}\,,
\end{equation}
and $p_{\hat \theta}= p_{\hat \phi}= w_2 \rho = \left(1+ \frac{v_c \Sigma(r, \theta) }{2} \right) \rho$.
For the energy density localized, we should set $v_c > 0$.
The energy density is positive/negative when $v_2 \gtrless 0$.
Therefore, we restrict our interest to the case $v_2 > 0$, satisfying the positive energy condition.

Because $v_1$ modifies the mass only, we set $v_1 = 0$ so that the  ADM mass of the black hole becomes just $M$.
Then,
\begin{eqnarray}
\label{deltasol}
&& \Delta(r) = r^2 - 2M r
+ a^2 +v_2 \bar \Delta(\sqrt{v_c} r);~~~~
\bar \Delta(y) \equiv e^{-y^2/2}
	+ \sqrt{\frac{\pi y^2}{2}} \left( {\rm erf}(\frac{|y|}{\sqrt{2}} )  -1\right) \,, \\
&& g_{tt} = -\left(1 - \frac{2Mr-v_2 \bar\Delta(\sqrt{v_c} r)}{r^2+a^2\cos^2\theta} \right)   \nonumber \,,
\end{eqnarray}
where $y = \sqrt{v_c} r $.

We now determine the locations of the ergosphere and the event horizon of the rotating black hole.
The ergosphere is the region located between the static limit surface and the outer event horizon~\cite{Ruffini:1970sp}.
The static limit surface is determined at the location $g_{tt}=0$, while
the event horizon corresponds to a Killing horizon~\cite{Carter:1969zz}, which is located where $\Delta(r)|_{r_H}=0$.
Due to the presence of a transcendental error function, precise determination of the horizon's position becomes challenging.
Therefore, we will numerically investigate it by assigning values to the coefficients $v_2$ and $v_c$,
as well as the rotation parameter $a$.

Before proceeding with the numerical analysis of the event horizon,
it is necessary to examine the constraints on the coefficient $v_2$ required for the event horizon to exist.
The function $\bar \Delta(y)$ monotonically decreases from one at $y=0$ to zero as $y\to \infty$.
Expanding $\Delta (r)$ around $r\sim 0$ and $r \gg v_c, M$, we get
\begin{equation}
\Delta (r) \approx a^2 + v_2 - \left(2M+ \sqrt{\frac{\pi}{2}}  v_2\right) r + \cdots, \qquad
\Delta (r) \approx r^2 - 2M r+ a^2 + O(r^{-\infty}) \,. \nonumber
\end{equation}
Here,
\begin{equation}
\Delta'(r) = 2(r-M) + v_2 \sqrt{\frac{\pi v_c}{2}}  \left( {\rm erf}(\sqrt{\frac{v_cr^2}{2} } )  -1\right) \,. \nonumber
\end{equation}
Let us set the radius $r_c$ such that $\Delta'(r_c ) = 0$ .
At this point, we have
\begin{equation}
\Delta(r_c) = a^2 - r_c^2 + v_2 e^{-v_c r_c^2/2} \,. \nonumber
\end{equation}
For an outer horizon to exist, the value of $v_2$ should be constrained by  the condition $ v_2< (r_c^2 -a^2) e^{ v_c r_c^2/2} $.
\begin{figure}[H]
\begin{center}
\subfigure[Horizon for varying the value of $v_2$.]
{\includegraphics[width=3.0in]{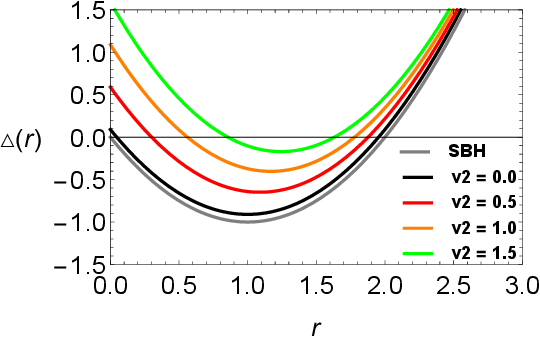}}
~~
\subfigure[Horizon for varying the value of $a$.]
{\includegraphics[width=3.0in]{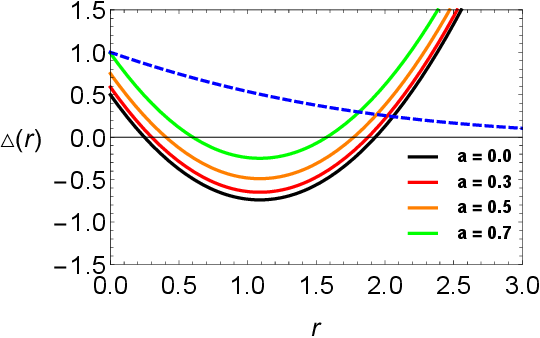}}
\end{center}
\caption{\footnotesize{(color online). The characteristic form of the function $\Delta(r)$ for various values of $v_2$ and $a$, respectively.}}
\label{hori}
\end{figure}

Figure \ref{hori} illustrates the shape of $\Delta(r)$ as the function of the coordinate $r$.
Figure \ref{hori}(a) shows the shape of $\Delta(r)$ while varying the value of $v_2$,
and Fig.~\ref{hori}(b) shows the shape while varying the value of the rotation parameter $a$.
We take $M=1$ for simplicity.
In Fig.~\ref{hori}(a), the gray curve represents a Schwarzschild black hole with $a=0$ and $v_2=0$, while the black curve represents a Kerr black hole with $v_2=0$.
The red, orange, and green curves represent $\Delta(r)$ with $v_2=0.5$, $v_2=1.0$, and $v_2=1.5$, respectively, with $a=0.3$.
The size of the event horizon is largest for a Schwarzschild black hole.
The second largest case is the outer event horizon of a rotating Kerr black hole.
When the rotation parameter $a$ held fixed, increasing $v_2$ leads to a reduction in the horizon size,
bringing the black hole closer to the extremal limit.
In Fig.~\ref{hori}(b), we take $v_2=0.5$. The blue dashed line represents $\bar \Delta(\sqrt{v_c}r)$ with $v_c=0.2$.
The black curve represents $\Delta(r)$ with $a=0.0$, corresponding to the black hole with  anisotropic matter.
The red, orange, and green curves represent $\Delta(r)$ with $a=0.3$, $a=0.5$, and $a=0.6$, respectively.
When the rotation parameter $a$ is set to zero, the outer horizon reaches its maximum size.
As $a$ increases, the horizon size shrinks and approaches the extremal limit.

Because of the existence of the matter field one can make extremal black hole and a naked singularity before reaching $a = M$.
To prevent it, we should restrict the range of $a$ to
\begin{equation}
v_2 \leq (r_c^2 -a^2) e^{ v_c r_c^2/2} \quad \rightarrow \quad
- \sqrt{ \frac{v_2^2 e^{-v_cr_c^2}}{4} + r_c^2} \leq a^2
	\leq  \sqrt{ \frac{v_2^2 e^{-v_cr_c^2}}{4} + r_c^2} \,. \nonumber
\end{equation}
Note that the presence of the $v_2 \bar \Delta(y) $ term always increases the value of $\Delta(r)$.
Therefore, the radius of the outer horizon $r_+$ (Henceforth, $r_H$ will be used to denote $r_+$.),
given by $\Delta(r)|_{r_H} =0$, decreases because of this term. This implies that the $\bar \Delta(y)$ term makes the entropy of the black hole decreases,
and as a result, the Kerr black hole maximizes the entropy.

Given the asymptotic values the mass $M$ and the angular momentum parameter $a$,  the horizon area varies with $v_c$.
As a result of this, the thermodynamic law can no longer be described by the asymptotic values.

In the $a \to 0$ limit, the metric reduces to Eq.~\eqref{gen Sch} with $\Delta^s(r)=r^2-2Mr+v_2 \bar \Delta(y)$, as given in Eq.~\eqref{deltasol},
this corresponds to a new static black hole geometry with $w_r=-1$, and $w_{\theta}=w_{\phi} =1+ v_c r^2/2 $.
In other words,
\begin{equation}
\rho^s = \frac{v_2 e^{-v_c r^2/2}}{8\pi r^4} \,,~~ p^s_{\hat r} = -\rho^s \,,~~
p^s_{\hat \theta}=p^s_{\hat \phi} = \left(1+ \frac{v_c r^2}2\right)\rho^s \,.
\label{eqofst}
\end{equation}
The energy density of the new matter field is similar to that of an electric field, $\rho_{c} = \frac{Q^2}{8\pi r^4}$,
but it has the property that its numerator decays exponentially as a function of the square of $r$,
rather than as a constant. The radial pressure equals the negative energy density,
resulting in a matter field that can coexist with the black hole outside the event horizon.
The transverse pressures have an additional contribution from the presence of $v_c$.
\begin{figure}[H]
\begin{center}
{\includegraphics[width=3.0in]{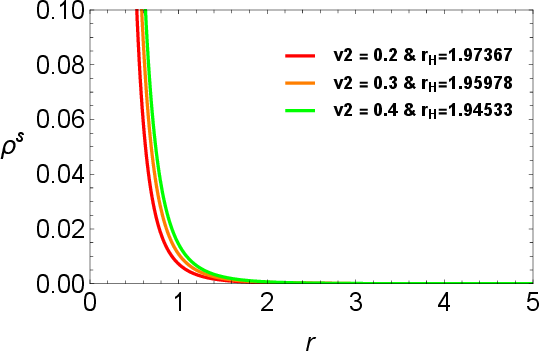}}
\end{center}
\caption{\footnotesize{(color online). The energy density as a function of $r$.}}
\label{density}
\end{figure}
Figure \ref{density} illustrates the profile of $\rho^s(r)$ as a function of $r$,
with parameters $M=1$ and $v_c=0.2$.
For the Schwarzschild black hole case, the horizon is located at $r_H=2$.
For black holes accompanying matter fields, however,
the presence of the error function complicates the precise determination of the event horizon.
To address this, we identify the horizon as the point where $\Delta^s(r)$ vanishes,
numerically approximated to within $10^{-5}$.
The red, orange, and green curves represent $\rho^s(r)$ with $v_2=0.2$, $v_2=0.3$, $v_2=0.4$ respectively.
As $v_2$ representing the contribution of matter field increases, the position of the outer horizon clearly shifts inward.

Now, let us turn to the mass energy. We begin with the case of a charged black hole.
The mass energy contained within a sphere of radius $R > r_H$ is
$m_c(R) = M- \frac{Q^2}{2R}$~\cite{Virbhadra:1990vs}.
The mass energy enclosed within the horizon, as viewed from the outside, is related to the irreducible mass $M_I$:
$m_c(r_H) = M -\frac{Q^2}{2r_H} = \frac{r_H}{2} \equiv M_I$, which implies $M=M_I + \frac{Q^2}{2r_H}$.
The second term corresponds to the contribution from the charge $Q$~~\cite{Christodoulou:1971pcn}.
In summary, the energy contained within the horizon is equivalent to the portion of
a black hole's total mass energy that cannot be extracted from the black hole.
For the present static black hole, The mass energy within the horizon is
\begin{equation}
\label{energy}
m(r_H)= M- \frac{v(r_H)}{2 r_H} \,,
\end{equation}
where $v(r_H)= v_2 \left[ e^{-\frac{r^2_H v_c}{2}} + \sqrt{\frac{\pi v_c}{2}} r_H \left( \mbox{erf} \left( \sqrt{\frac{v_c r^2_H}{2}} \right) -1 \right) \right] $.  This expression has the same structure as in the charged black hole case,
namely $m(r_H)=\frac{r_H}{2} \equiv M_I$, so that $M=M_I + \frac{v(r_H)}{2 r_H}$.
For the present rotating black hole, the mass energy relation from the horizon information of the black hole is given by
\begin{equation}
\label{irremass}
M^2 = \left[M_{Ir} + \frac{v(r_H)}{4M_{Ir}} \right]^2 + \frac{J^2}{4M^{2}_{Ir}}  \,,
\end{equation}
where $J=Ma$ denotes the angular momentum and $M_{Ir} \equiv \frac{\sqrt{r^2_H +a^2}}{2}$ denotes the irreducible mass of the rotating black hole.
For more information on other types of anisotropic matter fields, see~\cite{Kim:2019hfp}.

\section{Thermodynamics \label{sec4}}
\quad
We now turn to the thermodynamic properties of this black hole.
The surface gravity, which corresponds to the gravitational acceleration on the event horizon as measured by an observer at an asymptotic infinity, is a crucial quantity in black hole thermodynamics.
It is constant across the horizon and is directly proportional to the black hole temperature.
The black hole entropy has a geometrical interpretation, being proportional to the area of the black hole's horizon.
These thermodynamic quantities, as established in~\cite{Bekenstein:1973ur, Hawking:1975vcx}, are given by
\begin{equation}
\label{temp}
T_H=\frac{r^2_H -a^2 -v_2 e^{-v_c r^2_H/2} }{4\pi r_H (r^2_H + a^2)}\,,~~~ S=\pi (r^2_H +a^2)
\end{equation}
where $T_H$ represents the Hawking temperature and $S$ denotes the entropy.

\begin{figure}[H]
\begin{center}
\subfigure[Temperature for varying the value of $v_2$.]
{\includegraphics[width=3.0in]{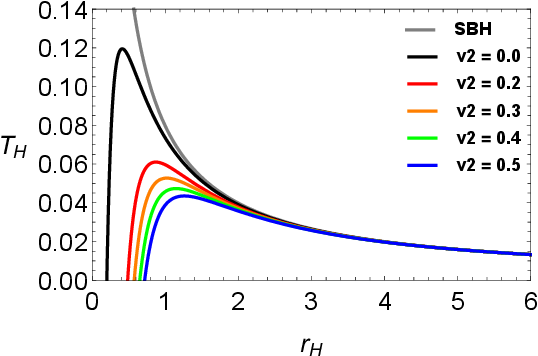}}
~~
\subfigure[Temperature for varying the value of $a$.]
{\includegraphics[width=3.0in]{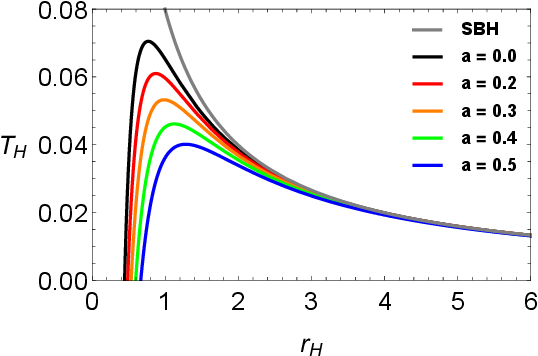}}
\end{center}
\caption{\footnotesize{(color online). Temperature as a function of the horizon radius $r_H$.}}
\label{temp}
\end{figure}
Figure \ref{temp} illustrates the black hole temperature as a function of the horizon radius  $r_H$ for the corresponding black hole solutions.
Figure \ref{temp}(a) shows temperature behavior with varying values of $v_2$, while Fig.~\ref{temp}(b) examines the effect of  varying the rotation parameter $a$.
We set $M=1$ for simplicity.
In Fig.~\ref{temp}(a), the grey curve represents the temperature of a Schwarzschild black hole $(a=0,v_2=0)$, while
the black curve corresponds to the temperature of a Kerr black hole $(a=0.2,v_2=0)$.
The red, orange, green, and blue curves depict the temperature profiles for $v_2=0.2$, $v_2=0.3$, $v_2=0.4$, and $v_2=0.5$, respectively.
As  $v_2$ increases, the maximum of the temperature shifts to the right and decreases in magnitude.
In Fig.~(b), the grey curve again corresponds to the temperature of a Schwarzschild black hole  $(a=0,v_2=0)$, while
the black curve represents the temperature of a black hole with anisotropic matter $(a=0, v_2=0.2)$.
The red, orange, green, and blue curves show the temperature profiles for $a=0.2$, $a=0.3$, $a=0.4$, and $a=0.5$, respectively.
As $a$ increases, the maximum temperature similarly shifts to larger $r_H$ and decreases in magnitude.

We now analyze the shape of the temperature curve for $T_H \geq 0$.
At large $r_H \gg (a^2+v_2e^{-v_c r^2_H/2})$, the temperature behaves as $T_H \propto r^{-1}_H$, similar to the Schwarzschild black hole.
Conversely, for small $r_H$, the term $-(a^2+v_2e^{-v_c r^2_H/2})$  in the numerator dominates.
In this regime, the temperature initially decreases as $T_H \propto -(r^{-3}_H)$ transitioning to  $T_H \propto -(r^{-1}_H)$ as $r_H$ grows slightly larger.
As $r_H \to 0$, $T_H$ diverges to negative infinity.
There exists a specific value of $r_H$ where $T_H=0$, corresponding to an extremal black hole.
Additionally, the $r_H$ value at which the black hole temperature reaches its maximum is determined by the condition $dT_H/dr_H=0$.

Black holes possess hairs~\cite{Ruffini:1971bza, Bardeen:1973gs}, such as  the mass, angular momentum, and the charge measurable by an asymptotic observer.
Using the Smarr relation~\cite{Smarr:1972kt}, we construct the Arnowitt-Deser-Misner mass.
The potential for the new field is determined by first deriving the  mass formula of a static, spherically symmetric black hole:
\begin{equation}
M=2T_H S + \Phi_{1} C_{1}+ \Phi_{2} C_{2} \,,
\label{masfors}
\end{equation}
where $C_{1}=\sqrt{v_2} v^{1/4}_c$,  $C_{2}=\sqrt{v_2}$, $\Phi_{1}=\sqrt{v_2} v^{1/4}_c \sqrt{\frac{\pi}{8}} \left( {\rm erf}\left( \sqrt{\frac{v_c}{2}} r_H \right) -1  \right)$, and $\Phi_{2}=\frac{\sqrt{v_2}}{r_H} e^{-\frac{v_c r^2_H}{2}}$ \,.
The new charges $C_1$ and $C_2$, however, cannot be measured asymptotically because the energy density~\eqref{eqofst} decreases exponentially with $r^2$.
The Smarr mass of the rotating black hole~\eqref{generalized Kerr} is given by
\begin{eqnarray}
M &&= \frac{1}{2}(2T_H S + 2 \Omega_H J  + \Phi_1 C_1+\Phi_2 C_2   )  \nonumber \\
   && + \frac{1}{2} \sqrt{(2T_H S + 2 \Omega_H J  + \Phi_1 C_1+\Phi_2 C_2  )^2-4\Omega_H J ( 2 \Phi_1 C_1+\Phi_2 C_2 )} \,,
\label{masforr}
\end{eqnarray}
where $\Omega_H=\frac{a}{r^2_H + a^2}$, $J=Ma$, $S=\pi(r^2_H + a^2)$. When $v_2 \to 0$,
it reduces to the case of the Kerr black hole with $M_K = 2T_H S + 2 \Omega_H J$.

We derive the first law of black hole thermodynamics to establish the differential relationship between the mass, the entropy,
the angular momentum, and the new charges of the black hole. The first law takes the form
\begin{eqnarray}
\delta M = \frac{M^2-\Omega_H  J \Phi_{1} C_{1} }{M^2} \left\{ T_H \delta S + \left(1- \frac{\Phi_1 C_1}{M} \right) \Omega_H \delta J + \left(1-\frac{\Omega_H J}{M} \right)
	\left[ 2\Phi_{1} \delta C_{1} + \Phi_{2} \delta C_{2} \right] \right\}\,.
\end{eqnarray}
This relation reduces to the Kerr black hole case, $\delta M_K = T_H \delta S + \Omega_H \delta J $ when $v_2 \to 0$.

Next, we analyze the specific heat (heat capacity) to evaluate
the thermodynamic local stability of the rotating black hole.
The heat capacity, $C=T_H \frac{\partial S}{\partial T_H}$,
is obtained as follows:
\begin{eqnarray}
C= \frac{2\pi r^2_H (r^2_H + a^2)
	[r^2_H - a^2
		-v_2 e^{-\frac{v_c r^2_H}{2}}
	]}
{ a^4 +4a^2r^2_H -r_H^4
	 +  [v_c r^2_H(r^2_H + a^2) + 3r^2_H + a^2]
	 	v_2 e^{-\frac{v_c r^2_H}{2}}
	 	} \,.
\end{eqnarray}
Notice that the numerator is nonnegative when $T_H \geq 0$.
Therefore, in this case, the signature of the heat capacity is determined by the value of the denominator.
When $v_2 \to 0$, this expression simplifies to that of the Kerr black hole:
\begin{eqnarray}
C_K= \frac{2\pi r^2_H (r^2_H + a^2)(r^2_H - a^2)}{a^4 + 4a^2r^2_H -r^4_H} \,.
\end{eqnarray}
When $a \to 0$, this reduces to that of the static case with corresponding the matter field as follows:
\begin{eqnarray}
C_{\rm gs} = \frac{2\pi r^2_H
	(r^2_H -v_2 e^{-\frac{v_c r^2_H}{2}})}
{ -r_H^2 + v_2 (v_c r^2_H + 3)e^{-\frac{v_c r^2_H}{2}}
	 	} \,. \nonumber
\end{eqnarray}
Please compare this result with that of the Schwarzschild.

\begin{figure}[H]
\begin{center}
\subfigure[Heat capacity for varying the value of $v_2$.]
{\includegraphics[width=3.0in]{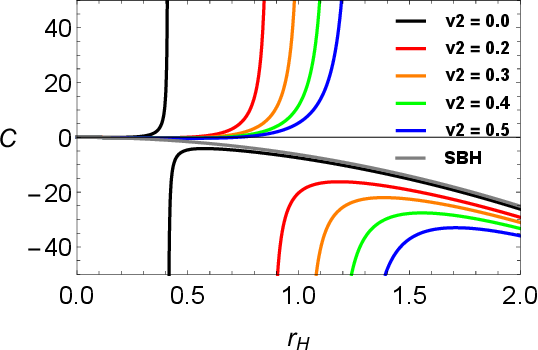}}
~~
\subfigure[Heat capacity for varying the value of $a$.]
{\includegraphics[width=3.0in]{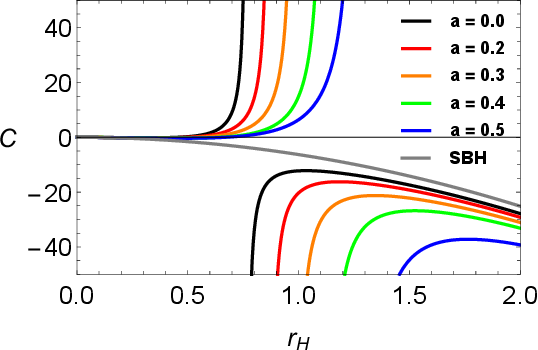}}
\end{center}
\caption{\footnotesize{(color online). Heat capacity as a function of the horizon radius $r_H$.}}
\label{heca}
\end{figure}
Figure~\ref{heca} illustrates the heat capacity as a function of the horizon radius  $r_H$ for the black hole solutions.
The black hole is locally thermodynamically stable when the heat capacity is positive and unstable when it is negative.
The singular point in each Figure corresponds to the convex upward point of the respective curve in Fig.~\ref{temp}, where $|\partial r_H/\partial T_H|\rightarrow \infty$.
Figure~\ref{heca}(a) depicts the heat capacity for varying value of $v_2$, while Fig.~\ref{heca}(b) examines the effect of different rotation parameter $a$ values.
In Fig.~\ref{heca}(a), the grey curve represents the heat capacity of a Schwarzschild black hole $(a=0,v_2=0)$.
The black curve corresponds to the heat capacity of the Kerr black hole with $(a=0.2,v_2=0)$.
The red, orange, green, and blue curves represent the heat capacity with $v_2=0.2$, $v_2=0.3$, $v_2=0.4$, and $v_2=0.5$, respectively.
In Fig.~\ref{heca}(b), the grey curve again corresponds to the Schwarzschild black hole.
The black curve represents a black hole with the anisotropic matter with $(a=0,v_2=0.2)$.
The red, orange, green, and blue curves illustrate the heat capacity for $a=0.2$, $a=0.3$, $a=0.4$, and $a=0.5$, respectively.

\section{Summary and discussions}

\quad

We used the results of the Newman-Janis (NJ) algorithm~\cite{Newman:1965tw}
and the ``non-complexification procedure''~\cite{Azreg-Ainou:2014pra}
as a method of separation of variables in the metric to solve the Einstein equation
and obtain the reduced form of the metric~\eqref{metric:ortho gen}.
Then, we presented a new solution to the Einstein equations describing a rotating black hole
as an extension of the Kerr spacetime.
We first demand the Einstein tensor component  $R_{r \theta}$ vanishes so that the geometry is consistent with the axial symmetry.
We then, check the Einstein tensor to get proper equation of states  the stress tensor should follow.

However, obtaining a rotating black hole solution becomes more challenging
when the NJ algorithm is not well-defined for a given geometry.
This issue arises, for instance, when $f(r) \neq g(r)$, when a cosmological constant is present, or when the black hole is immersed in a magnetic field.
In the case of $f(r) \neq g(r)$, the standard approach is to solve Einstein's equations directly by using a metric ansatz,
and verify whether $R_{r \theta}=0$.
An example of this method applied within the NJ algorithm framework can be found in~\cite{Kim:2024mam}.
For black holes with a cosmological constant, one may follow Carter's method~\cite{Carter:1973rla},
while black holes immersed in the magnetic field require analysis based on Ernst spacetime~\cite{Ernst:1976mzr, Ernst:1976bsr}.

The energy budget of our Universe is composed of approximately $5\% $ ordinary matter
and $95\%$ dark energy and dark matter~\cite{Planck:2018nkj}.
Recent astrophysical and cosmological precision observations have intensified the trend toward studying cosmological and astrophysical models involving dark energy and dark matter~\cite{Ko:2016dxa, Lee:2019ums, Biswas:2023eju, Lee:2025zol}.
These phenomena have motivated us to study black holes that coexist with matter fields,
with the goal of understanding and explaining their nature.

Notably, we found that the geometry of the rotating black hole solution is not uniquely determined by asymptotic parameters such as mass, charge, angular momentum, or any other gauge charges. Specifically, a matter field exists near the black hole horizon, with its density decaying not later than exponentially at large distances.
This solution, therefore, challenges the no-hair conjecture and suggests the existence of a new form of black hole``hair" that remains undetectable in the asymptotic region.
Investigating the properties of such a matter field in curved spacetime would be an intriguing direction for future research.
An alternative, indirect approach is to analyze black hole thermodynamics by defining a new charge associated with this field and identifying the corresponding potential.

We analyzed black hole thermodynamics, defining the charge for the new field and identifying the corresponding potential.
This allowed us to derive the Smarr relation and the first law of black hole thermodynamics.
This approach indirectly demonstrates the existence of the new hair.
Additionally, we analyzed the black hole temperature and heat capacity, which reveals the thermodynamic local stability of the black hole.

\section*{Acknowledgments}
We are grateful to Wontae Kim and Stefano Scopel for their hospitality
during our visit to the Workshop on Cosmology and Quantum Spacetime (CQUeST 2024), and Inyong Cho to the
workshop on How to use AI in Astrophysics Theory.
H.-C.~Kim (RS-2023-00208047) and W.~Lee (RS-2022-NR075087, CQUeST: RS-2020-NR049598)
were supported by Basic Science Research Program through
the National Research Foundation of Korea funded by the Ministry of Education.
We thank Metin Gurses for giving valuable comments.

\end{document}